\documentclass[showpacs,twocolumn,preprintnumbers,amsmath,amssymb,nofootinbib]{revtex4}
\usepackage{graphicx}
\usepackage{dcolumn}
\usepackage{bm}

\begin{document}
\title{\bf TDIR: Time-Delay Interferometric Ranging \\
for Space-Borne Gravitational-Wave Detectors}

\author{Massimo Tinto}
\email{Massimo.Tinto@jpl.nasa.gov}
\altaffiliation [Also at: ]{Space Radiation Laboratory, California
  Institute of Technology, Pasadena, CA 91125}
\affiliation{Jet Propulsion Laboratory, California Institute of Technology, Pasadena, CA 91109}

\author{Michele Vallisneri}
\email{Michele.Vallisneri@jpl.nasa.gov}
\affiliation{Jet Propulsion Laboratory, California Institute of Technology, Pasadena, CA 91109}

\author{J. W. Armstrong}
\email{John.W.Armstrong@jpl.nasa.gov}
\affiliation{Jet Propulsion Laboratory, California Institute of Technology,
 Pasadena, CA 91109}

\date{\today} 
\begin{abstract}
  Space-borne interferometric gravitational-wave detectors, sensitive
  in the low-frequency (mHz) band, will fly in the next decade.  In
  these detectors, the spacecraft-to-spacecraft light-travel times
  will necessarily be unequal and time-varying, and (because of
  aberration) will have different values on up- and down-links.  In
  such unequal-armlength interferometers, laser phase noise will be
  canceled by taking linear combinations of the laser-phase
  observables measured between pairs of spacecraft, appropriately
  time-shifted by the light propagation times along the corresponding
  arms.  This procedure, known as time-delay interferometry (TDI),
  requires an accurate knowledge of the light-time delays as functions
  of time. Here we propose a high-accuracy technique to estimate these
  time delays and study its use in the context of the Laser
  Interferometer Space Antenna (LISA) mission. We refer to this
  ranging technique, which relies on the TDI combinations themselves,
  as Time-Delay Interferometric Ranging (TDIR). For every TDI
  combination, we show that, by minimizing the rms power in that
  combination (averaged over integration times $\sim \, 10^4$ s) with
  respect to the time-delay parameters, we obtain estimates of the
  time delays accurate enough to cancel laser noise to a level well
  below the secondary noises. Thus TDIR allows the implementation of
  TDI without the use of dedicated inter-spacecraft ranging systems,
  with a potential simplification of the LISA design.  In this paper
  we define the TDIR procedure formally, and we characterize its
  expected performance via simulations with the \textit{Synthetic
    LISA} software package.
\end{abstract}

\maketitle


The Laser Interferometer Space Antenna (LISA) \cite{benderetal} is a
planned NASA--ESA mission to detect and study gravitational waves
(GWs) in the $10^{-4}$--1 Hz band, by exchanging coherent laser beams
between three widely separated spacecraft. Each spacecraft contains
two drag-free proof masses, which provide freely falling references
for the GW-modulated relative spacecraft distances; these are measured
by comparing the phases of the incoming and local lasers. It has been
shown that the time series of the phase differences can be combined,
with suitable time delays, to cancel the otherwise overwhelming laser
phase noise, while preserving a GW response. This technique is known
as time-delay interferometry (TDI; see \cite{STEA03,TEA2004} and
references therein).

In the case of a stationary LISA spacecraft array, it was estimated
\cite{TSSA2003} that the time delays need to be known with an accuracy
of about $100$ ns, if the various TDI combinations are to work
effectively, suppressing the residual laser phase fluctuations to a
level below the secondary noises (such as the proof-mass and
optical-path noises).  For an array of spacecraft in relative motion
along realistic Solar orbits, more complicated
(\emph{second-generation}) TDI combinations are needed; these require
an even more accurate knowledge of the time delays \cite{VTA04}. The
most direct implementation of TDI consists in triggering the phase
measurements at the correct delayed times (within the required
accuracy), as suggested in Ref.\ \cite{TSSA2003}. This approach
requires the real-time, onboard knowledge of the light-travel times
between pairs of spacecraft, which determine the TDI time delays.
Although the triggering approach is feasible in principle, it
complicates the design of the optical phasemeter system, and it
requires an independent onboard ranging capability. Recently, it was
pointed out \cite{SWSV2004} that the phase measurements at the
specific times needed by the TDI algorithm can be computed \emph{in
  post-processing} with the required accuracy, by the fractional-delay
interpolation (FDI) \cite{SWSV2004,laakso} of regularly sampled data
(with a sampling rate of 10 Hz for a GW measurement band extending to
1 Hz).

In this communication, we show that FDI allows the implementation of a
numerical variational procedure to determine the TDI time delays from
the phase-difference measurements themselves, eliminating the need for
an independent onboard ranging capability. Since this variational
procedure relies on the TDI combinations, we refer to it as Time-Delay
Interferometric Ranging (TDIR).


Conventional spacecraft ranging is based on the measurement of either
one-way or two-way delay times.  In one-way ranging, two or more tones
are coherently modulated onto the transmitted carrier; the phases of
these tones are measured at the receiver, differenced, and divided by
the spanned bandwidth to yield the group delay and hence the time
delay (up to an ambiguity of $c$ divided by the spanned bandwidth of
the ranging tones).  In two-way ranging, a known ranging code is
modulated on the transmitted carrier, which is transponded by a
distant spacecraft back to the originator; the received signal is then
cross-correlated with the ranging code to determine the two-way time
of flight.

TDIR differs from these methods in that it uses the unmodulated laser
noises in a three-element array, which are canceled in TDI
combinations assembled with the correct inter-spacecraft light-travel
times. This means that TDI can be used to estimate the light-travel
times by minimizing the laser noise power in the TDI combinations as a
function of the postulated light-travel times: this process defines
TDIR. As an example of how TDIR works, we shall here consider one of
the second-generation TDI combinations, the unequal-armlength
Michelson combination $X_1$ \cite{TEA2004},
\begin{widetext}
\begin{multline}
X_1 = \left[(\eta_{31} + \eta_{13;\hat2}) + (\eta_{21} + 
  \eta_{12;\hat3'})_{;\hat2'\hat2} + (\eta_{21} + \eta_{12;\hat3'})_{;\hat3\hat3'\hat2'\hat2} + (\eta_{31} +
  \eta_{13;\hat2})_{;\hat3\hat3'\hat3\hat3'\hat2'\hat2} \right] \\
-  \left[(\eta_{21} + \eta_{12;\hat3'}) + (\eta_{31} +
  \eta_{13;\hat2})_{;\hat3\hat3'} + (\eta_{31} + \eta_{13;\hat2})_{;\hat2'\hat2\hat3\hat3'} + 
(\eta_{21} + \eta_{12;\hat3'})_{;\hat2'\hat2\hat2'\hat2\hat3\hat3'} \right].
\label{eq:1}
\end{multline}
\end{widetext}
Here we use the notation of Ref.\ \cite{TEA2004}, where the
$\eta_{ij}$ (for spacecraft indices $i,j = 1, 2, 3, \ i \ne j$) are
linear combinations of the inter-spacecraft phase measurements
$s_{ij}$ and of the inter-bench measurements $\tau_{ij}$ made aboard
the spacecraft,
\begin{equation}
\label{eq:2}
\begin{aligned}
\eta_{21} & \equiv  s_{21} - {1 \over 2} \ [\tau_{32} - \tau_{12}]_{;\hat 3'}\,,
& \eta_{31} & \equiv s_{31} + {1 \over 2} \ [\tau_{21} - \tau_{31}]\,, \\
\eta_{12} & \equiv s_{12} + {1 \over 2} \ [\tau_{32} - \tau_{12}]\,, 
& \eta_{32} & \equiv s_{32} - {1 \over 2} \ [\tau_{13} - \tau_{23}]_{;\hat 1'}\,, \\
\eta_{13} & \equiv s_{13} -  {1 \over 2} \ [\tau_{21} - \tau_{31}]_{;\hat 2'}\,, 
& \eta_{23} & \equiv s_{23} + {1 \over 2} \ [\tau_{13} - \tau_{23}]\,,
\end{aligned} 
\end{equation}
and where indices prefixed by a semicolon delay observables by the
corresponding light-travel times $L_k$, in the sequential order given
(with the index $k=1$ denoting the light travel time for the beam
emitted from spacecraft 2 and received at spacecraft 3, and likewise
indices $\{2, 3, 1', 2', 3'\} \equiv \{3 \rightarrow 1,1 \rightarrow
2,3 \rightarrow 2,1 \rightarrow 3,2 \rightarrow 1\}$).

The main contributions to the phase measurements $s_{ij}$ and
$\tau_{ij}$ are given by
\begin{equation}
\label{eq:3}
\begin{aligned}
s_{31} & = \phi_{13;2} - \phi_{31} + \Sigma_{31} \, , \\
s_{21} & = \phi_{12;3'} - \phi_{21} + \Sigma_{21} \, , \\
\tau_{31} & = \phi_{21} - \phi_{31} + \Lambda_{31} \, , \\
\tau_{21} & = \phi_{31} - \phi_{21} + \Lambda_{21}
\end{aligned}
\end{equation}
(and by cyclical permutations thereof), where the $\phi_{ij}$ denote
the sum of laser phase fluctuations and of optical-bench motions (the
former three to four orders of magnitude larger than the latter), and
where the $\Sigma_{ij}$ and the $\Lambda_{ij}$ denote the sum of all
other fluctuations affecting the measurements, such as the secondary
noise sources (proof mass and optical path) and GWs.

The time-delay indices that appear in Eq.\ \eqref{eq:3} represent the
actual delays caused by the physical propagation of the laser signals
across the LISA arms. By contrast, the hatted delays of Eq.\ 
\eqref{eq:1} need to be provided by the data analyst (or, in the
triggering approach, by the onboard ranging subsystem) with the
accuracy required for effective laser noise cancellation.  Thus, the
$X_1$-based implementation of TDIR works by \emph{minimizing the power
  in $X_1$ with respect to the hatted delays $\hat{L}_k$}. Since the
TDI combinations constructed with the actual delays cancel laser phase
noise to a level $10^8$ below the secondary noises \cite{STEA03}, it
follows that if we neglect all non-laser sources of phase noise (i.e.,
if we set $\Sigma_{ij} = \Lambda_{ij} = 0$), the minimum of the power
integral
\begin{equation}
I^{(0)}(\hat{L}_k) = \frac{1}{T} \,\int_0^T [X^{(0)}_{1}(\hat{L}_k)]^2 \, dt
\label{LSQ0}
\end{equation}
will occur for $\hat{L}_k = L_k$ (with $k = 1,2,3,1',2',3'$; here the
superscript ${}^{(0)}$ denotes \emph{laser-noise--only} quantities).
The search for this minimum can be implemented in post processing,
using FDI \cite{SWSV2004} to generate the needed $s_{ij}$ and
$\tau_{ij}$ samples at the delayed times corresponding to any choice
of the $\hat{L}_k$.

In reality, the presence of non-laser phase noises (possibly including
GWs) will displace the location of the minimum from $L_k$. Writing
$X_1 = X_1^{(0)} + X_1^{(n)}$ (with $X_1^{(n)}$ obtained by setting
all $\phi_{ij} = 0$), the power integral becomes
\begin{equation}
I^{(n)}(\hat{L}_k) = \frac{1}{T} \, \int_0^T [X_{1}(\hat{L}_k)]^2 \, dt \,,
\label{LSQ1}
\end{equation}
or explicitly,
\begin{equation}
\begin{aligned}
I^{(n)}(\hat{L}_k) = I^{(0)}(\hat{L}_k) &+ \frac{1}{T} \, \int_0^T [X_1^{(n)}]^2 \, dt \\
& + \frac{2}{T} \, \int_0^T X_1^{(n)} X_1^{(0)}(\hat{L}_k) \, dt \, .
\end{aligned}
\label{LSQ2}
\end{equation}
Here we have written the non-laser phase noise $X_1^{(n)}$ as
independent of the delays $\hat{L}_k$: this holds true for a search
conducted sufficiently close to the true minimum, since the
$\phi_{ij}$ are much larger than the $\Sigma_{ij}$ and $\Lambda_{ij}$,
and so are their variations. The minimum of $I^{(n)}(\hat{L}_k)$ can
be displaced from $\hat{L}_k = L_k$ because the third term of Eq.\ 
\eqref{LSQ1} [the cross-correlation integral of $X_1^{(n)}$ and
$X_1^{(0)}(\hat{L}_k)$] can be negative and offset a concurrent
increase in $I^{(0)}(\hat{L}_k)$.  The achievable time-delay
accuracies will depend on the level of the residual laser noise, the
levels of the secondary noises in $X_1$, and the integration time $T$.
We expect the arm-length errors to be determined by the interplay of
the first and third terms in Eq.\ \eqref{LSQ2}.  By equating the
variance from the imperfect cancellation of the laser with the
estimation-error variance of the cross-term in Eq.\ \eqref{LSQ2}, we
can roughly estimate how well the time delays will be determined with
TDIR: $\delta L_k \sim
(\sigma_{X_{1}^{(n)}}/{\sigma_{\dot{X}_{1}^{(0)}}}) \ 
\sqrt{{\rho}/{T}}$, where $\sigma_{X_{1}^{(n)}}$ and
$\sigma_{\dot{X}_{1}^{(0)}}$ are the root-mean-squares of the
secondary noises and of the time derivative of the laser noise in
$X_1$, and $\rho$ is the temporal width of the secondary-noise
autocorrelation function.  For nominal LISA noises and $T \simeq$
10,000 s we thus expect $\delta L_k$ of 30 ns or better to be
achievable.

An analogous technique was proposed by G\"ursel and Tinto \cite{GT89}
for the problem of determining the parameters of a GW burst observed
in coincidence by a network of ground-based GW interferometers.  In
that case, a ``phase-closure'' condition was imposed on a family of
linear combinations of the responses of three GW detectors.  A GW
burst would produce a zero response in the particular phase-closed
combination corresponding to the source's position in the sky; the
position could then be estimated by implementing a least-squares
minimization procedure. In that case as well as for TDIR, the
least-squares minimization procedure can be shown to be optimal if the
secondary noises have Gaussian distribution.


We test TDIR for a realistic model of the LISA orbits and instruments
by performing simulations with the \emph{Synthetic LISA} software
package \cite{MV04}. Because the present version of \emph{Synthetic
  LISA} works in terms of frequencies rather than phases, we perform
an analogue of the procedure outlined above where all the phase
variables are replaced by the corresponding fractional-frequency
fluctuations.  We generate a number of \emph{chunks} of contiguous
data for the $s_{ij}$ and $\tau_{ij}$ measurements, sampled at
intervals of 0.25 s, and containing pseudo-random laser, proof-mass,
and optical-path noises at the nominal level set by the LISA pre-phase
A specification \cite{benderetal,MV04}. We consider chunk durations of
8,192, 16,384, and 32,768 s.

The 18 noise processes (corresponding to the six lasers, proof masses,
and optical paths) are assumed to be uncorrelated, Gaussian, and
stationary, with (respectively) white, $f^{-2}$, and $f^{2}$ PSDs,
band-limited at 1 Hz. The frequency-fluctuation measurements contain
also the responses due to GWs from two circular binaries with
$f_\mathrm{GW} \simeq$ 1 and 3 mHz, located respectively at the vernal
equinox and at ecliptic latitude $45^\circ$ and longitude $120^\circ$.
The strength of the two sources is adjusted to yield an optimal S/N of
$\sim$ 500 over a year (for $X_1$), guaranteeing that there will be
times of the year when each source will be clearly visible above the
noise in an observation time $\sim$ 10,000 s.

We put the three LISA spacecraft on realistic trajectories, modeled as
eccentric, inclined solar orbits with angular velocity $\Omega =
2\pi/$yr, average radius $R/c \simeq 499$ s, and eccentricity $e
\simeq 9.6 \times 10^{-3}$ \cite{CR03}.  The resulting time and
direction dependence \cite{STEA03} of the light travel times is then
\cite{CR03,MV04}
\begin{equation}
\begin{aligned}
L_{k}(t)
= L & + \frac{1}{32} (e L) \sin(3 \Omega t - 3 \xi_0) \\
& -  [\frac{15}{32}(e L) \pm (\Omega R L)] \sin(\Omega t - \delta_k)\,,
\end{aligned}
\label{eq:lfunc}
\end{equation}
where the plus (minus) refers to unprimed (primed) indices. In Eq.\ 
\eqref{eq:lfunc} $L/c \simeq 16.68$ s is the average light travel
time, and
\begin{equation}
(\delta_1,\delta_2,\delta_3) =
(\xi_0,\xi_0+\frac{4\pi}{3},\xi_0+\frac{2\pi}{3})\,,
\end{equation}
with $\xi_0$ an arbitrary constant (set to 0 in our simulations)
giving the phase of the spacecraft motion around the guiding center of
the LISA array. The starting times of the chunks are spread across a
year to sample the time dependence of the $L_k$ and the directionality
of the GW responses.

\begin{figure*}
\includegraphics[width=6.4in]{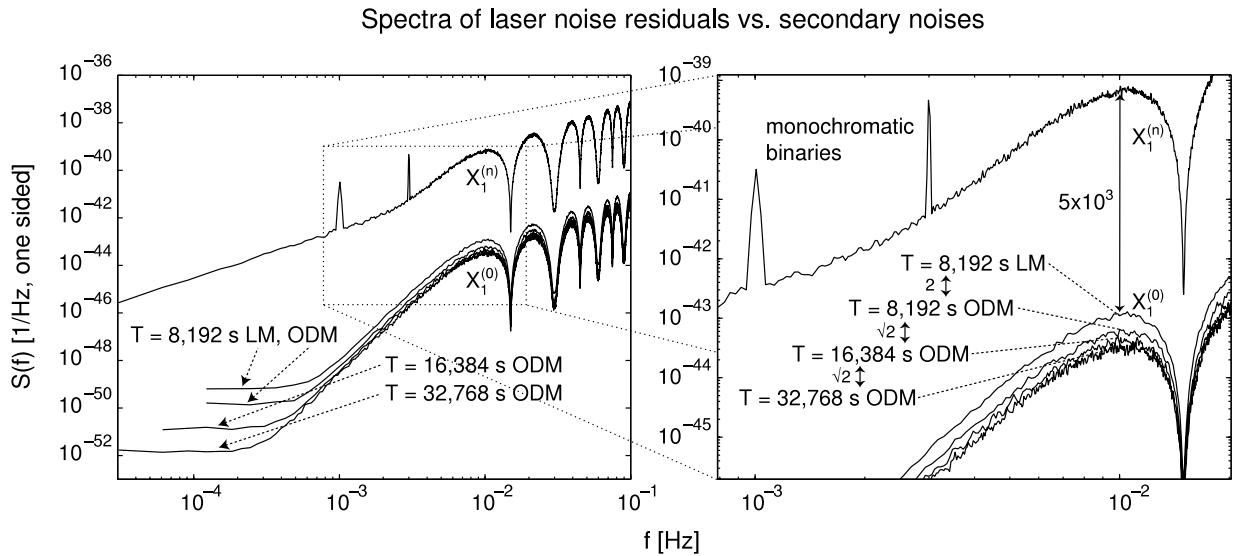}
\caption{Spectra of frequency laser noise (bottom curves) and of GW 
plus secondary noises (top curve) at the end of TDIR minimization 
using chunk durations of 8,192 s (for the LM and ODM models), and 
16,384 and 32,768 s (for the ODM model only). We show averages of 
the spectra computed separately for each chunk using a
triangle-windowed periodogram; the averages are taken over 
populations of of 512, 256, and 128 chunks for $T = $ 8,192, 
16,384, and 32,768 s respectively, spread across a year. In all cases, 
laser noise is suppressed to levels several orders of magnitude below 
the secondary noises: the cutout graph on the right shows that the 
typical laser-noise suppression factor with respect to secondary 
noise is $\sim 5 \times 10^3$ for the worst case considered 
(8,192-s LM); it improves by a factor $\sim 2$ for 8,192-s ODM, 
and by factors of $\sim \sqrt{2}$ for each successive doubling 
of $T$. The GWs from the two circular binaries stand clearly 
above the noise at 1 and 3 mHz. 
\label{fig:spectra}}
\end{figure*}

Separately for each chunk, we minimize $I^{(n)}[\hat{L}_k(t)]$ [Eq.\ 
\eqref{LSQ1}] starting from guesses for the $\hat{L}_k$ affected by
errors $\gtrsim$ 50 $\mathrm{km}/c$, very much larger than typical
accuracy of radio tracking from Earth \cite{Folkner}. The minimization
is carried out using a Nelder--Mead simplex-based algorithm \cite{NM}.
The effective cancellation of laser noise with TDI requires modeling
the time dependence of the travel times \emph{within} the chunks. In
our simulations, we use two such models:
\begin{enumerate}
\item An orbital-dynamics model (ODM) given by Eq.\ \eqref{eq:lfunc}, with
  $\widehat{eL}$, $\widehat{\Omega R L}$, and $\widehat{\xi_0}$ taken
  as the independent search parameters with respect to which $I^{(n)}$
  is minimized. We exclude $L$ and $\Omega$ from the search because
  the dependence of the $L_k(t)$ on such an extended parameter set is
  degenerate on time-scales $\sim$ 10,000 s.
\item A linear model (LM) given by $\hat{L}_k(t) = \hat{L}^0_k +
  \hat{L}^1_k (t - t_0)$ [with $t_0$ set to the beginning of each
  chunk]. Because the expression for $X_1$ does not contain the travel
  times $L_1$ and $L_{1'}$, our independent search parameters are the
  constants $\hat{L}^0_k$ and $\hat{L}^1_k$ for $k = 2, 2', 3, 3'$
  (eight numbers altogether).
\end{enumerate}
\begin{figure}
\includegraphics[width=3.2in]{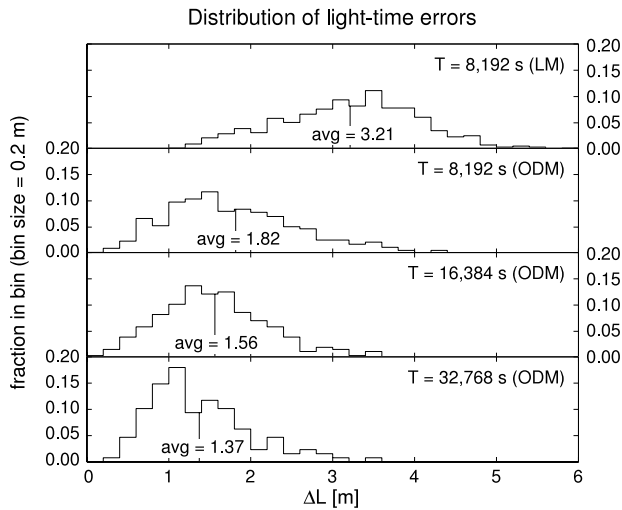}
\caption{Distribution of errors $\Delta L$ [see Eq.\ \eqref{eq:errL}
and the main text above it] in the determination of light travel
times, using $X_1$-based TDIR with chunk durations of 8,192 s 
(for the LM and ODM models), and 16,384 and 32,768 s (for the ODM 
model only). As expected, the errors are lower for longer integration 
times $T$; for the LM model, the larger errors are due to the
unmodeled curvature in the time dependence of the light-travel times. 
The distributions shown correspond to samples of 512, 256, and 128 
chunks for $T = $ 8,192, 16,384, and 32,768 s respectively, spread 
across a year.
\label{fig:errs}}
\end{figure}
Figures \ref{fig:errs} and \ref{fig:spectra} show the results of our
simulations. The average travel-time errors $\Delta L$ displayed in
Fig.\ \ref{fig:errs} are defined as $\Delta L = (\Delta L_2 + \Delta
L_{2'} + \Delta L_3 + \Delta L_{3'})/4$, with
\begin{equation}
\Delta L_k = \sqrt{\frac{1}{T} \int_{t_0}^{t_0+T}
\left( \hat{L}_k(t) - L_k(t) \right)^2 \, dt}.
\label{eq:errL}
\end{equation}
Because the noises have different realizations in each chunk and
because the local behavior of the $L_k(t)$ [Eq.\ \eqref{eq:lfunc}]
changes along the year, the average error $\Delta L$ of each chunk is
a random variable. Its distribution is approximated by the histograms
of Fig.\ \ref{fig:errs}, which refer to populations of respectively
512 (for T = 8,192 s), 256 (for T = 16,384 s), and 128 (for T = 32,768
s) chunks (hence the roughness of the curves).

It turns out that the linear model is not quite sufficient to model
the changes of the time-delays during the chunk lengths considered,
since the \emph{minimum} $\Delta L$s [computed by least-squares
fitting the parameters $\hat{L}^0_k$ and $\hat{L}^1_k$ to the
$L_k(t)$] are in the range 0.25--2.60 m (for T = 8,192 s), 1--10 m
(for T = 16,384 s), and 4--40 m (for T = 32,768 s). Thus, in Figs.\ 
\ref{fig:errs} and \ref{fig:spectra} we show results only for the
linear model with T = 8,192 s. [The minimization of $I^{(n)}$ over the
LM parameters is delicate, because for $X_1$ the laser-noise residuals
turn out to depend strongly on $\Delta L_2$, $\Delta L_{3'}$, and
$\Delta L_{2'} - \Delta L_3$, but only weakly on $\Delta L_{2'} +
\Delta L_3$. In this case, the Nelder--Mead algorithm can be made to
return accurate results by using the search parameters $\hat{L}^0_2$,
$\hat{L}^0_{3'}$, $\widehat{L^0_{2'}-L^0_{3}}$, and
$\widehat{L^0_{2'}+L^0_{3}}$, plus the corresponding $\hat{L}^1_k$
parameters.]

Figure \ref{fig:spectra} shows the spectra of the \emph{residual}
laser noise [i.e., of $X^{(0)}_1$ at the minimum of
$I^{(n)}(\hat{L}_k)$], as compared with spectra of GWs and secondary
noises [i.e., of $X^{(n)}_1$]. The spectra are computed separately for
each chunk using triangle-windowed periodograms, and then averaged
over the chunk populations. The two GW sources stand clearly above the
secondary noises at 1 and 3 mHz.  We see that the TDI cancellation of
laser noise with TDIR-determined time-delays is essentially complete,
with the residual laser noise several orders of magnitude below the
secondary noises.  We conclude that for $T \sim 10,000$ s, with the
nominal LISA noises, and even in the presence of very strong GW
signals, TDIR can easily reach the time-delay accuracy required for
second-generation TDI.  For frequencies below 10 mHz, the residual
laser-noise power decays as $f^6$, while the secondary noises decrease
only as $f^2$. We attribute the flattening near 0.1 mHz (which is
insignificant with respect to the LISA performance) to a combination
of leakage and aliasing in the numerical estimation of the spectra and
of real effects due to the first non-constant terms in the travel time
errors across the chunks.

Finally, we estimate the power in the Fourier bins containing the
simulated signals using two different time series: in the first $X_1$
was formed using perfectly known time delays, in the second using the
TDIR-determined time delays. Analyzing the 32,768-s chunks at the
times along the simulated year where the signal amplitudes were
maximum, we find that the signal powers in the two time series agree
to the numerical precision of the calculation (about a part in
$10^5$).


In summary, we propose a method that uses TDI and the intrinsic phase
noise of the lasers in a three-element array to determine the
inter-spacecraft light travel times. This method, Time-Delay
Interferometric Ranging (TDIR), relies on the fact that TDI nulls all
the laser noises when the time delays are chosen to match the travel
times experienced by the laser beams as they propagate along the sides
of the array. Simulations performed using the nominal LISA noises
indicate that, for integration times $\sim 10,000$ s, TDIR determines
the time delays with accuracies sufficient to suppress the laser phase
fluctuations to a level below the LISA secondary noises, while at the
same time preserving GW signals. Our simulations assume synchronized
clocks aboard the spacecraft, but we anticipate that TDIR may be
extended to achieve synchronization, by minimizing noise power also
with respect to clock parameters.

TDIR has the potential of simplifying the LISA design, allowing the
implementation of TDI without a separate inter-spacecraft ranging
subsystem. At the very least, TDIR can supplement such a subsystem,
allowing the synthesis of TDI combinations during ranging dropouts or
glitches. TDIR may be applicable in other forthcoming space science
missions that rely on spacecraft formation flying and on
inter-spacecraft ranging measurements to achieve their science
objectives.

The TDIR technique presented in this paper was based on the TDI
combination $X_1$. Since the accuracy in the estimation of the time
delays achievable by TDIR depends on the magnitudes of the secondary
noises entering the specific TDI combination used, it is clear that it
should be possible to optimize the effectiveness of TDIR over the
space of TDI combinations \cite{PTLA2002}.
We will investigate this problem in a more extensive future article.

\acknowledgments 
We thank F.\ B.\ Estabrook for his long-time
collaboration and for many useful conversations during the development
of this work. M.\ V.\ was supported by the LISA Mission Science Office
at the Jet Propulsion Laboratory. This research was performed at the
Jet Propulsion Laboratory, California Institute of Technology, under
contract with the National Aeronautics and Space Administration.


\begin{references}
%
\bibitem{benderetal} P. L. Bender, K. Danzmann, and the LISA Study Team, \emph{Laser Interferometer Space Antenna for the Detection of
Gravitational Waves, Pre-Phase A Report}, 2nd ed.,
doc. MPQ 233 (Max-Planck-Instit\"ut f\"ur Quantenoptik, Garching, Germany, 1998).
%
\bibitem{STEA03} D. A. Shaddock, M. Tinto, F. B. Estabrook, and J. W. Armstrong, Phys. Rev. D {\bf 68}, 061303(R) (2003). 
%
\bibitem{TEA2004} M. Tinto, F.B. Estabrook, and J.W. Armstrong, Phys. Rev. D {\bf 69}, 082001 (2004).
%
\bibitem{TSSA2003} M. Tinto, D. A. Shaddock, J. Sylvestre, and J. W. Armstrong, Phys. Rev. D {\bf 67}, 122003 (2003).
%
\bibitem{VTA04} M. Vallisneri, M. Tinto, and J. W. Armstrong, in preparation (2004).
%
\bibitem{SWSV2004} D. A. Shaddock, B. Ware, R. E. Spero, and M. Vallisneri,
Phys. Rev. D, in print (2004); gr-qc/0406106.
%
\bibitem{laakso} T. I. Laakso et al., IEEE Signal Processing Magazine {\bf 13}, 30 (1996).
%
\bibitem{JenkinsWatts} G. M. Jenkins and D. G. Watts, \emph{Spectral Analysis and Its Applications} (Holden-Day, San Francisco, 1968).
%
\bibitem{GT89} Y. G\"ursel and M. Tinto, Phys. Rev. D {\bf 40}, 3884 (1989).
%
\bibitem{MV04} M. Vallisneri, gr-qc/0407102.
%
\bibitem{CR03} N. Cornish and L. Rubbo, Phys. Rev. D {\bf 67}, 022001 (2003).
%
\bibitem{Folkner} W. Folkner, private communication (2004).
%
\bibitem{NM} J. A. Nelder and R. Mead, Computer Journal \textbf{7}, 308 (1965).
%
\bibitem{PTLA2002} T. A. Prince, M. Tinto, S. L. Larson, and J. W. Armstrong, Phys. Rev. D \textbf{66}, 122002 (2002).
%
\end{references}
\end{document}